\documentclass[preprint,amsmath,amssymb,aps,showkeys,showpacs,pra]{revtex4-1}

\usepackage[english]{babel}
\usepackage{graphicx}
\usepackage{graphics}
\usepackage{amsmath}
\usepackage{dcolumn}
\usepackage{amssymb}
\usepackage{bm}
\usepackage[latin1]{inputenc}
\usepackage{graphicx,color}

\begin{document}
\title{Evolution of Landau levels in graphene-based topological insulators in the presence of wedge disclinations}

\author{J.R.S. Oliveira}
\email{}
\affiliation{Departamento de Fisica, Universidade Federal da Para\'{i}ba, Caixa Postal 5008, 58051-970  Jo$\tilde{a}$o Pessoa, PB, Brazil}

\author{G.Q. Garcia}
\email{}
\affiliation{Departamento de Fisica, Universidade Federal da Para\'{i}ba, Caixa Postal 5008, 58051-970  Jo$\tilde{a}$o Pessoa, PB, Brazil}

\author{C. Furtado}
\email{Corresponding author, furtado@fisica.ufpb.br} 
\affiliation{Departamento de  Fisica, Universidade Federal da Para\'{i}ba, Caixa Postal 5008, 58051-970  Jo$\tilde{a}$o Pessoa, PB, Brazil} 

\author{S. Sergeenkov}
\email{} 
\affiliation{Departamento de  Fisica, Universidade Federal da Para\'{i}ba, Caixa Postal 5008, 58051-970  Jo$\tilde{a}$o Pessoa, PB, Brazil}

\begin{abstract}
In this paper  we consider modification of electronic properties of graphene-based topological insulator in the presence of wedge disclination and magnetic field by adopting the Kane-Mele model with intrinsic spin-orbit coupling.  Using the properly defined Dirac-Weyl equation for this system,  an exact solution for the  Landau levels is obtained.  The influence of the topological defect on the evolution of Landau levels is discussed.
\end{abstract}


\maketitle

\section{Introduction}
The physics of graphene still remains a rich area both for fundamental research and promising  applications \cite{gein,been,castro}. Recently, Kane and Mele \cite{prl:kane} have investigated the role of the intrinsic ($\Delta$) and Rashba type ($\lambda_{R}$) spin-orbit coupling (SOC) in a single graphene layer. In particular, they found that accounting for SOC converts an ideal two-dimensional semi-metallic state of graphene into a quantum spin Hall insulator. Later, this new class of electronic materials, gapped in the bulk but with topologically protected edge states near the boundary of the sample \cite{RMP:kane}, was coined \textit{topological insulators}. Using angle-resolved photoemission spectroscopy (ARPES) technique, Pan {\it et al.} \cite{pan} have confirmed manifestation of topological insulator properties in $Bi_2Se_3$ compound. Similar results were obtained by Roushan {\it et al.} \cite{rous} for $Bi_{1-x}Sb_x$ alloys which indicated the presence of a large energy gap and single-surface Dirac cone  associated with topologically protected state in this material \cite{dubo}. 
Regarding theoretical studies in this area, it is worth mentioning a recent work by De Martino {\it et al.} \cite{demart} who have analysed the modification of the electronic properties in graphene monolayer in the presence of an applied magnetic field and pseudomagnetic field created by intrinsic and Rashba type SOC contributions. The study of the Landau levels in graphene layer with Rashba coupling revealed \cite{rash} the emergence of two zero modes energy states. A low-energy spectrum of the Landau levels in bilayer graphene in the presence of transverse magnetic field and a Rashba based SOC was investigated by Mireles and Schliemann \cite{mire}. The structure of the Landau levels for a series of gapped Dirac materials (such as silicene, germanene, etc)  with intrinsic Rashba interaction was investigated by Tsaran and Sharapov \cite{tsar}. Some interesting properties of the bound states spectrum emerging in graphene-based topological insulator have been obtained by De Martino {\it et al.} \cite{cris} within the Kane-Mele Hamiltonian in the presence of attractive potential and considering both intrinsic and Rashba type SOC. In recent years, the influence of topological defects on electronic properties of graphene has been widely investigated by several authors ~\cite{rmp:voz,epl:voz,prl:lam,pla:car,KnuteClaudio1,knut,carvalho}. For example,  R\"uegg and Lin ~\cite{rueg} have studied the bound states induced by a disclination  in  graphene-based  topological insulators. By using the  Haldane honeycomb lattice model \cite{hald} for spherical on spherical nano surfaces (tetrahedron, octahedron and icosahedron), R\"uegg {\it et al.} ~\cite{rueg:prb}  concluded  that each corner of these structures (named \textit{topological fullerenes}) forms a non-trivial bound state.  Choudhari and  Deo \cite{epl:niv} considered the influence of disclination on electronic properties of a single graphene layer with intrinsic and Rashba type SOC and found the energy spectrum for a modified Kane-Mele Hamiltonian. 

In this paper we use the modified Kane-Mele Hamiltonian for disclinated single-layer graphene with intrinsic SOC contribution in the presence of topological defect induced magnetic flux and a uniform magnetic field in order to study the eigenfunctions and eigenvalues  of   Kane-Mele Hamiltonian for disclinated graphene layer  - the Landau levels in this geometry. Some physical implications of the found results will be discussed.

\section{Kane-Mele model for disclinated single-layer graphene}\label{sec2}
In this Section we analyse the low-energy electronic properties of disclinated graphene layer within the Kane-Mele model in the presence of an intrinsic SOC contribution. 
Recall that such systems can be reasonably treated introducing a multi-dimensional tight-binding bases. To describe graphene, the tight-binding Hamiltonian is used which allows for hopping of electrons between nearest neighbors  (from one sublattice to another) in a manner that electrons on an atom of the type $A/B$ can hop on the three nearest $B/A$ atoms. By taking into account the hopping between next-to-nearest neighbors (introduced by Haldane \cite{haldane}), we can write down the resulting Kane-Mele \cite{prl:kane} Hamiltonian as follows:
\begin{eqnarray}
\mathcal{H} = t\sum_{\alpha}\sum_{<ij>}c^{\dagger}_{i \alpha} c_{j \alpha} + it_{2}\sum_{\alpha \beta}\sum_{<<ij>>} \nu_{ij}s^{z}_{\alpha \beta} c^{\dagger}_{i \alpha} c_{j \beta}
\end{eqnarray}
Here, $t$ and $t_{2}$ stand for the nearest and next-to-nearest neighbors hopping amplitudes, respectively.
Notice that the second term describes the connection between second neighbors with a spin dependent amplitude.  $\nu_{ij}$ depends upon the direction of second nearest neighbours hopping (anticlockwise is positive, clockwise is negative). $s^{z}_{\alpha \beta}$ is the spin operator for an electron. 

The low-energy continuous limit of the above tight-binding Hamiltonian with intrinsic spin-orbit coupling (SOC) is given by:
\begin{eqnarray}
\mathcal{H} =\hbar v_{f}(\tau_{z}\sigma_{x} k_{x} +  \sigma_{y} k_{y}) + \triangle  \tau_{z}\sigma_{z}s_{z},
\end{eqnarray} 
where $\tau_{i}$, $\sigma_{i}$, and $s_{i}$ are Pauli matrices acting on states in valleys,  sublattices, and spin spaces, respectively. We use the following notation: $\tau_0$,  $s_0$ are the identities for Pauli matrices, $\tau_{z}=\pm 1$ for two valleys $K(K')$ in Brillouin zone,  $\sigma_{z}=\pm 1$ for sublattices $A(B)$, and $s_{z}=\pm 1$ for up and down electron spins.  $\triangle   = 3 \sqrt{3} t_{2}$ is the SOC parameter for the honeycomb lattice which produces a gap in the energy spectrum. 
The Hamiltonian $\mathcal{H}$ acts on the eight-component spinor 
$\Psi= [(\psi_{A\uparrow}\psi_{A\downarrow}\psi_{B\uparrow}\psi_{B\downarrow}),(\psi_{A'\uparrow}\psi_{A'\downarrow}\psi_{B'\uparrow}\psi_{B'\downarrow})]^{T}$, where $A$ and $B$ label the sublattices in the valley $K$, and $A'$ and $B'$  label the sublattices in the valley $K'$.

The low-energy continuum limit for treating the graphene layer with disclination can be described by geometric theory of defects \cite{bilbi1,bilbi2,kroner,volovik}. It is based on the metric that contains all information about the elastic deformations caused by the disclination  \cite{katanaev}. This approach is similar to  geometric theory of gravity with curvature and torsion. 
The conical geometry can be presented in various ways \cite{jacdese}. For a sake of comparison, in this paper we discuss three different representations.  The first representation is the geometric  representation  of coordinates that describe a cone imbedded in flat $3D$-space with the constraint $z=\sqrt{(\alpha_{N}^{2}-1)(x^{2} + y^{2})}$ where $(x,y,z)$ are the usual Euclidean coordinates and $\alpha_{N}=\sin(\beta)$ with $\beta$ being the aperture angle of the cone. This geometry is described by the  following metric 
\begin{eqnarray}\label{imbeddedmetric}
ds^{2}=\frac{d\rho^{2}}{\alpha_{N}^{2}} +\rho^{2}d \varphi^{2},
\end{eqnarray}
where $0<\rho < \infty$ and  $0<\varphi<2\pi$. 

It is important to point out that in the present paper we have used  geometric description of conical space \cite{jacdese} related to the "unfolded plane" concept \cite{rueg,epl:niv} which is based on intrinsic characterization and uses an Euclidean geometry with incomplete angular range  given by the following metric 
\begin{equation}\label{metricplane}
ds^{2}= dr^{2} + r^{2}d\phi^{2},
\end{equation}
where $0<r<\infty$ and $0<\phi<2\pi\alpha_{N}$ with  $\alpha_{N}$ being the intensity of  the disclination, which can be written in terms of the angular sector $\lambda$ (which we removed or inserted in the graphene layer to form the defect)  as  $\alpha =1 \pm \lambda / 2 \pi$. This geometric description is the second representation of conical geometry. Recall that graphene layer can be described as a hexagonal lattice with two carbon atoms in the unitary cell  \cite{castro}. A  pictorial manner to view the introduction of a topological defect into graphene lattice is the Volterra process \cite{volterra}.  This cut and glue process  obeys the symmetry of the honeycomb lattice and the sector $\lambda$  is a multiple of $\pi /3$,  so that $\lambda=\pm N \pi /3$ where  $N$  is  an  integer   with $0<N< 6$.  When the  parameter $\alpha_{N}$ is within the range $ 0 <\alpha_{N}<1$, it characterizes  a positive disclination, where  within the  Volterra process we remove an angular sector $- N \pi /3$ of graphene layer. When we insert an angular sector $ N \pi /3$ we obtain negative disclination. For example, the insertion of  $\lambda=  \pi /3$ results in a heptagon in apex of  disclinated medium.  Notice that may exist  saddle-shape cones in graphene for which the value of $N \pi /3$  or $\alpha_{N}>1$. This negative disclinated medium has negative curvature and the insertion of this sector  introduces saddle point that characterizes this conical space of negative curvature. Good discussion about negative disclination and its geometry can be found in  \cite{sitenko}.  In  a graphene layer, disclinations are described by the line  element~(\ref{metricplane})  corresponding to removed sector (positive disclination) or inserted sector (negative disclination). 
The third representation describes a cone embedded in three-dimensional flat space and is given by the metric
\begin{equation}\label{metriccone}
ds^{2}= dr^{2} + {\alpha_{N}}^{2}r^{2}d\theta^{2},
\end{equation}
where $0<r<\infty$ and $0<\theta<2\pi$.   The connection between the above three representations is realized through the relations $\rho =\alpha_{N}r$ and $\varphi=\alpha_{N}^{-1}\phi$  which connects the representations (\ref{imbeddedmetric})  and (\ref{metricplane}) according to Fig.1, and the relation $\theta=\frac{\phi}{\alpha_{N}}$ which links the representations (\ref{metricplane}) and (\ref{metriccone})  \cite{jacdese,premisha,rueg,epl:niv}.  

\begin{figure}
\includegraphics[scale=0.5]{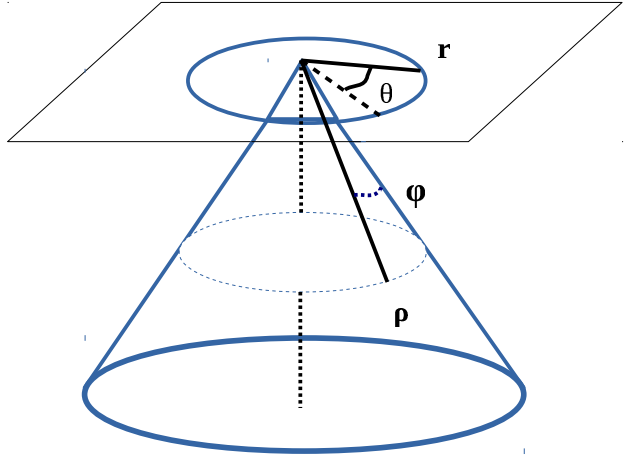} 
\caption{(Color online) The representation of geometries of imbedded cone and the polar representation. The coordinates of representations (\ref{imbeddedmetric}) and (\ref{metricplane}) are shown.}
\label{fig:fig1}
\end{figure}

The topological defects are known to affect the electronic properties of graphene \cite{ser1,ser2,ser3,ser4}. In particular, such defect can be introduced via quantum fluxes of fictitious gauge fields through the apex of the conical space. In this way the transport parallel to the wave function with spin around disclination  \cite{prl:lam,prb:lam,pla:car,pr:katn,np:cor,rueg,epl:niv}  produces a non-trivial holonomy. Thus, to include the mismatch of the phases of the spinor base wave functions, we have to perform a gauge transformation of the spinor  $\Psi=[(\psi_{A\uparrow}\psi_{A\downarrow}\psi_{B\uparrow}\psi_{B\downarrow}),(\psi_{A'\uparrow}\psi_{A'\downarrow}\psi_{B'\uparrow}\psi_{B'\downarrow})]^{T}$ due to the defect in the  graphene layer.  These phases introduced by fictitious gauge fields in the envelope function of spinor are responsible  for making the total eigenspinor single-valued.  The general  holonomy that describes these fluxes in honeycomb lattice with disclination is given by
\begin{eqnarray}\label{holonomy}
\psi(\theta=2\pi)=e^{i\frac{N\pi}{6}(\tau_{z}\sigma_{z}s_{0} - 3\tau_{y}\sigma_{y}s_{0})}\psi(\theta=0).
\end{eqnarray}
This holonomy manifests itself as a change of reference frames in space in the presence of topological defect and the consequent change of the spinor due to the mismatch of "cut and glue" process.
We can rewrite the expression for holonomy  (\ref{holonomy}) in two contributions.  One is given by $U_{s}(\phi)=e^{i\frac{\phi}{2}\tau_{z}\sigma_{z}s_{0}}$ and is due to the parallel transport of spinor  around the apex  of cone in a closed path, thus proving the variation of the local reference frame along the path ~\cite{prl:lam,pla:car,rueg}. As is pointed out in ~\cite{rueg,epl:niv}, $U_{s}(\phi)$ is responsible for $\Psi$ translation  in co-rotating  spinor. The other holonomy $V_{ns}(\theta)=e^{i\frac{N\theta}{4}\tau_{y}\sigma_{y}s_{0}}$ is of Aharonov-Bohm-type ~\cite{pr:ahan} contribution \cite{prl:lam} and it introduces a  matrix-valued gauge field in the  Hamiltonian of the system leading to
\begin{eqnarray}\label{transfo}
\psi(\theta=2\pi)=U_{s}(\phi)V_{Ns}(\theta)\psi(\theta=0).
\end{eqnarray}
Let us consider now the two external magnetic fields.  The first one is the Aharonov-Bohm flux ~\cite{pr:ahan}. Its appearance in conical space has been discussed in much detail in \cite{rueg,epl:niv}. The unfolded plane coordinates of the vector potential $\vec{A}$, that generates a magnetic flux in the center of disclination, is given by
\begin{eqnarray}\label{abflux}
\vec{A}=\frac{\Phi}{r\alpha_{N}\Phi_{0}} \hat{\phi}
\end{eqnarray}
Here,  $\Phi_{0}=h/2e$ is the magnetic flux quanta.

The second external  field is a uniform magnetic field $\vec{B} = B_{0}\hat{z}$ in conic space. The configuration that generates this potential in the presence of disclination was obtained in \cite{pla:bruno} and is given by 
\begin{eqnarray}\label{unifor}
\vec{A_{r}}= \frac{B_{0}r}{2}\hat{\phi}
\end{eqnarray}
For the electron with spin, using the Kane-Mele model at low energies in polar coordinates $(r,\theta)$ with these magnetic fields (\ref{abflux}) and (\ref{unifor}) but with the modified intrinsic SOC Hamiltonian $H'_{so}$ for the conical space, we can write down the resulting Kane-Mele Hamiltonian as follows
\begin{eqnarray}\label{kanemele}
H_{KM}=\hbar v_{f}\left[\tau_{z}\sigma_{x}  k_{r} +  \sigma_{y} \left(k_{\theta} + \frac{\Phi}{r\alpha_{N}\Phi_{0}} +  \frac{\pi B_{0}r}{\Phi_{0} }\right) \right] + H'_{so},
\end{eqnarray}
 where $k_{r}=-i\frac{\partial}{\partial r}$ and $k_{\theta}=\frac{-i}{r\alpha_{N}}\frac{\partial} {\partial\theta}$. The Hamiltonian $H'_{so}$ due to the SOC term in graphene with a disclination ~\cite{epl:niv} reads
\begin{eqnarray}\label{socdis}
H'_{so}=\triangle  [\tau_{z}\sigma_{z}s_{z}h(\beta)+ \tau_{z}\sigma_{z}s_{x}cos(\theta)p(\beta) - \tau_{z}\sigma_{z}s_{y}sin(\theta)p(\beta) ]
\end{eqnarray}
Here $h(\beta)=\left(1 - \frac{p^{2}}{4}\right)$ with $p(\beta)=\frac{2\cos\beta}{1+ \sin\beta}$  where $\beta$ is the  angle of aperture of the cone formed by the wedge disclination in the graphene layer. Here, we use the same condition that (\ref{socdis}) $\theta=2\pi$ due to the fact that $\theta=\frac{\phi}{1\pm \frac{N}{6} }$ and for $\phi =2\pi -\lambda$ we always have \cite{epl:niv} $\theta=2\pi$. This way we can rewrite the SOC term  (\ref{socdis}) as follows
\begin{eqnarray}\label{socdismod}
H'_{so}=\triangle  [\tau_{z}\sigma_{z}s_{z}h(\beta)+ \tau_{z}\sigma_{x}s_{x}p(\beta)]
\end{eqnarray}
Using the gauge transformation (\ref{transfo}) in the Hamiltonian (\ref{kanemele}), we have 
\begin{eqnarray}
  H_{D} =U_{s}^{\dagger}(\phi) V^{\dagger}_{ns}(\theta)H_{KM}U_{s}(\phi) V_{ns}(\theta) \nonumber
\end{eqnarray}  
The resulting  transformed Hamiltonian for disclinated graphene lattice is given by 
\begin{eqnarray}\label{disclanatedhamil}
H_{D} = \left[k_{r}- \frac{i}{2r} \right]\tau_{z}\sigma_{x}  + \left[k_{\theta}+ \frac{\Phi}{r\alpha_{N}\Phi_{0}} 
+  \frac{e B_{0}r}{2} + \frac{N}{4r\alpha_{N}} \right] \sigma_{y} +\nonumber \\ + \triangle  [\tau_{z}\sigma_{z}s_{z}h(\beta)+ \tau_{z}\sigma_{x}s_{x}p(\beta)] 
\end{eqnarray}
In what follows, we use the system of unities with $\hbar= v_{f}=1$. To find the energy spectrum based on the Hamiltonian (\ref{disclanatedhamil}), we apply the {\it Ansatz}:  $\psi(r,\theta)=e^{ij\theta}X(r)$, where $j$ is a half integer number ($j = \pm1/2, \pm3/2,...$), resulting in the following form of the radial part of Hamiltonian 
\begin{eqnarray}\label{modifiedkane}
H'_{D}(r)= \left(k_{r}-\frac{i}{2r} \right)\tau_{z}\sigma_{x}   + \left(\frac{\nu_{\tau}}{r} + \frac{eB_{0}r}{2} \right) \sigma_{y}  + \triangle  [\tau_{z}\sigma_{z}s_{z}h(\beta)+ \tau_{z}\sigma_{x}s_{x}p(\beta)]
\end{eqnarray}
Where $\tau=\pm$ for two emergent valleys, and $\nu_{\tau}$ is given by
\begin{eqnarray}
\nu_{\tau}=\frac{j+\frac{\Phi}{\Phi_{0}}+\frac{N\tau}{4}}{\alpha_{N}}
\end{eqnarray}

\section{Landau Levels for graphene-based topological insulator with a single disclination}\label{sec3}
In this Section using the modified Kane-Mele Hamiltonian in disclinated graphene we study the Landau levels  for graphene-based topological insulator. We consider Dirac-Weyl Hamiltonian with an intrinsic SOC. The role of a Rasbha type SOC contribution will be treated in a separate article.

Now, let us solve the Dirac-Weyl equation $H'_{D}X(r)=\epsilon X(r)$ for the modified Kane-Mele Hamiltonian  (\ref{modifiedkane}) to obtain the corresponding eigenvalues $\epsilon$ and eigenfunctions $X(r)$ for the seeking problem. We get four coupled equations for the first valley $K$ and another four equations for the second valley $K'$. More specifically, the set of coupled equations for the first valley reads 
\begin{eqnarray}
\label{acoplada1}p(\beta)\triangle X_{A\downarrow}(r) - i\left[\frac{d}{dr}+ \frac{1}{2r}+ \nu_{+}+ \omega r \right]X_{B\uparrow}(r)=(\epsilon - \triangle  h(\beta))X_{A\uparrow}(r),\\
\label{acoplada2}p(\beta)\triangle X_{A\uparrow}(r) - i\left[\frac{d}{dr}+ \frac{1}{2r}+ \nu_{+}+ \omega r \right]X_{B\downarrow}(r)=(\epsilon + \triangle  h(\beta))X_{A\downarrow}(r),\\
\label{acoplada3}-p(\beta)\triangle X_{B\downarrow}(r) - i\left[\frac{d}{dr}+ \frac{1}{2r}- \nu_{+} - \omega r \right]X_{A\uparrow}(r)=(\epsilon + \triangle  h(\beta))X_{B\uparrow}(r), \\
\label{acoplada4}-p(\beta)\triangle X_{B\uparrow}(r) - i\left[\frac{d}{dr}+ \frac{1}{2r}- \nu_{+} - \omega r \right]X_{A\downarrow}(r)=(\epsilon - \triangle  h(\beta))X_{B\downarrow}(r).
\end{eqnarray} 
In turn, for valley $K'$ we have the following set of equations
\begin{eqnarray}
\label{acoplada1'}-p(\beta)\triangle  X_{A'\downarrow}(r) + i\left[\frac{d}{dr}+ \frac{1}{2r}- \nu_{-}- \omega r \right]X_{B'\uparrow}(r)=(\epsilon + \triangle  h(\beta))X_{A'\uparrow}(r), \\
\label{acoplada2'}-p(\beta)\triangle  X_{A'\uparrow}(r) + i\left[\frac{d}{dr}+ \frac{1}{2r}- \nu_{-} - \omega r \right]X_{B'\downarrow}(r)=(\epsilon - \triangle  h(\beta))X_{A'\downarrow}(r), \\
\label{acoplada3'}p(\beta)\triangle  X_{B'\downarrow}(r) + i\left[\frac{d}{dr}+ \frac{1}{2r}+ \nu_{-} + \omega r \right]X_{A'\uparrow}(r)=(\epsilon - \triangle  h(\beta))X_{B'\uparrow}(r),  \\
\label{acoplada4'}p(\beta)\triangle  X_{B'\uparrow}(r) + i\left[\frac{d}{dr}+ \frac{1}{2r} + \nu_{-} + \omega r \right]X_{A'\downarrow}(r)=(\epsilon + \triangle  h(\beta))X_{B'\downarrow}(r),
\end{eqnarray} 
where we introduced the definition $\omega = \frac{eB_{0}}{2}$. If we multiply (\ref{acoplada1}) by $(\epsilon + \triangle  h)$ and use (\ref{acoplada2}) and (\ref{acoplada3}), we can find the equation for $X_{A\uparrow}$. Likewise, by multiplying (\ref{acoplada4}) by $(\epsilon + \triangle  h)$ and using (\ref{acoplada2}) and (\ref{acoplada1}), we can find the equation for $X_{B\downarrow}$. After decoupling these equations, we obtain for the first valley
\begin{eqnarray}
\frac{d^{2}X_{A\uparrow}}{dr^{2}}+ \frac{1}{r}\frac{dX_{A\uparrow}}{dr}-\left[\kappa^{2} + \frac{1}{r^{2}}\left(\nu_{+}-\frac{1}{2}\right)^{2}+ w_{+} + \omega^2 r^2 \right] X_{A\uparrow} = 0\\
\frac{d^{2}X_{B\downarrow}}{dr^{2}} + \frac{1}{r}\frac{dX_{B\downarrow}}{dr}-\left[\kappa^{2} + \frac{1}{r^{2}}\left(\nu_{+}+\frac{1}{2}\right)^{2}+ w'_{+} + \omega^2 r^2\right] X_{B\downarrow} = 0
\end{eqnarray}
Here, $\kappa^{2} = \sqrt{(h^{2}+p^{2})\triangle  ^{2} - \epsilon^{2})}$, $h(\beta)=\left( 1 - \frac{p^{2}}{4}\right)$, $ w_{+}=eB_{0}( \nu_{+} +\frac{1}{2} )$, $w'_{+}=e B_{0}(\nu_{+} -\frac{1}{2})$ and $\omega= \frac{eB_{0}}{2}$. 

After performing the coordinate transformation $\rho=\omega r^{2}$, the equations take this form
\begin{eqnarray}
\rho\frac{d^{2}X_{A\uparrow}(\rho)}{d\rho^{2}} + \frac{dX_{A\uparrow}(\rho)}{d\rho}- \left[\frac{1}{4\omega}(w_{+}+\kappa^2) + \frac{\rho}{4}+\frac{\zeta^{2}_{A\uparrow}}{4\rho}\right]X_{A\uparrow}=0\\
\rho \frac{d^{2}X_{B\downarrow}(\rho)}{d\rho^{2}} + \frac{d X_{B\downarrow}(\rho)}{ d\rho}- \left[\frac{1}{4\omega}(w'_{+}+\kappa^2) + \frac{\rho}{4}+\frac{\zeta^{2}_{B\downarrow}}{4\rho}\right]X_{B\downarrow}=0
\end{eqnarray}
To construct plausible and physically sound solutions of the above equations, we can use their asymptotic behavior at two critical points, $\rho \to 0 $ and $\rho \to \infty$. Our analysis results in the following choice for the radial eigenfunctions  
\begin{eqnarray}
X_{A\uparrow}(\rho)= e^{-\rho/2}\rho^{\frac{\vert\zeta_{A\uparrow}\vert}{2}}F_{A\uparrow}(\rho)\\
X_{B\downarrow}(\rho)=e^{-\rho/2}\rho^{\frac{\vert\zeta_{B\downarrow}\vert}{2}}F_{B\downarrow}(\rho)
\end{eqnarray}
The functions $F_{A(A')\uparrow(\downarrow)}(\rho)$ and $F_{B(B')\downarrow(\uparrow)}(\rho)$ satisfy  the Hypergeometric equations (or Kummer equations). It can be easily verified that similar results take place for the second valley $K'$ as well
\begin{eqnarray}
X_{A'\downarrow}(\rho)= e^{-\rho/2}\rho^{\frac{\vert\zeta_{A'\downarrow}\vert}{2}}F_{A'\downarrow}(\rho)\\
X_{B'\uparrow}(\rho)=e^{-\rho/2}\rho^{\frac{\vert\zeta_{B'\uparrow}\vert}{2}}F_{B'\uparrow}(\rho).
\end{eqnarray}
In the above solutions, we made use of the following definitions: 
\begin{eqnarray}
\left \vert\zeta_{A\uparrow}\right \vert =\left  \vert \nu_{+}-\frac{1}{2}\right \vert\\
\left \vert\zeta_{B\downarrow}\right \vert =\left  \vert \nu_{+}+\frac{1}{2}\right  \vert\\
\left \vert\zeta_{A'\downarrow}\right \vert =\left  \vert \nu_{-}+\frac{1}{2}\right \vert\\
\left \vert\zeta_{B'\uparrow}\right\vert = \left \vert \nu_{-}-\frac{1}{2}\right \vert
\end{eqnarray}
In this way, we obtain the following set of differential  Hypergeometric equations for $F_{A(A')\uparrow(\downarrow)}(\rho)$ and $F_{B(B')\downarrow(\uparrow)}(\rho)$
\begin{eqnarray}
&\rho&\frac{d^{2}F_{A\uparrow}}{d\rho ^{2}} +  \left[1 + |\zeta_{A\uparrow} | - \rho \right]  \frac{d F_{A\uparrow}}{d \rho} + \left[-\frac{|\zeta_{A\uparrow} |}{2} - \frac{1}{2} - \frac{1}{4\omega}\left(w_{+} + \kappa^{2}\right) \right]F_{A\uparrow}  = 0   \\
&\rho&\frac{d^{2}F_{A\downarrow}}{d\rho ^{2}}  +\left[1 + |\zeta_{B\downarrow} | - \rho \right]\frac{d F_{B\downarrow}}{d\rho}  + \left[-\frac{|\zeta_{B\downarrow} |}{2} - \frac{1}{2} - \frac{1}{4\omega}\left(w'_{+} + \kappa^{2}\right) \right]F_{B\downarrow}  = 0  \\
&\rho&\frac{d^{2}F_{A'\downarrow} }{d\rho^{2}}+  \left[1 + |\zeta_{A'\downarrow} | - \rho \right]\frac{d F_{A'\downarrow}}{d\rho} +  \left[-\frac{|\zeta_{A'\downarrow} |}{2} - \frac{1}{2} - \frac{1}{4\omega}\left(w'_{-} + \kappa^{2}\right) \right]F_{A'\downarrow} = 0   \\
&\rho&\frac{d^{2}F_{B'\uparrow}}{d\rho^{2}} +\left[1 + |\zeta_{B'\uparrow} | - \rho \right] \frac{d F_{B'\uparrow}}{d\rho}  +\left[-\frac{|\zeta_{B'\uparrow} |}{2} - \frac{1}{2} -\frac{1}{4\omega}\left(w_{-} + \kappa^{2}\right) \right]  F_{B'\uparrow} = 0 
\end{eqnarray}
where $w_{+}=eB_{0}( \nu_{+} +\frac{1}{2})$, $w'_{+}=\frac{e B_{0}}{\alpha_{N}}(\nu_{+} -\frac{1}{2})$, $w_{-}=\frac{eB_{0}}{\alpha_{N}}( \nu_{-} +\frac{1}{2})$,  $w'_{-}=\frac{e B_{0}}{\alpha_{N}}(\nu_{-} -\frac{1}{2})$. 

To avoid divergence of the obtained solutions at the critical points $\rho \to 0 $ and $\rho \to \infty$, we use the standard for Hypergeometric series $F(a,b,\rho)$ procedure by imposing the condition $a=-n$ where $n=0, 1,2,3,4 ...$. Using this  condition, we arrive at the following set of regularized equations
\begin{eqnarray}
F_{A\uparrow}(\rho) = F\left(\left[\frac{\vert \zeta_{A\uparrow}\vert}{2}+\frac{1}{2}+\frac{(w_{+}+\kappa^{2})}{4\omega} \right],\vert \zeta_{A\uparrow}\vert + 1; \rho \right), \\
F_{A'\downarrow}(\rho)= F\left(\left[\frac{\vert \zeta_{A'\downarrow}\vert}{2}+\frac{1}{2}+\frac{(w'_{-}+\kappa^{2})}{4\omega} \right],\vert \zeta_{A'\downarrow}\vert + 1; \rho \right),  \\
F_{B\downarrow}(\rho) = F\left(\left[\frac{\vert \zeta_{B\downarrow}\vert}{2}+\frac{1}{2}+\frac{(w'_{+}+\kappa^{2})}{4\omega} \right],\vert \zeta_{B\downarrow}\vert + 1; \rho \right), \\
F_{B'\uparrow}(\rho) = F\left(\left[\frac{\vert \zeta_{B'\uparrow}\vert}{2}+\frac{1}{2}+\frac{(w_{-}+\kappa^{2})}{4\omega} \right],\vert \zeta_{B'\uparrow}\vert + 1; \rho \right). 
\end{eqnarray}
As a final result of this paper, we obtain 
\begin{eqnarray}\label{spinorcom}
\psi(\rho,\theta) = C_{n,j}e^{ij\theta}\left[ \begin{array}{c}
e^{-\rho/2}\rho^{\frac{\vert\zeta_{A\uparrow}\vert}{2}}F_{A\uparrow}(\rho)\\ \nonumber \\
\frac{p\triangle  e^{-\rho /2}\rho^{\frac{\vert \zeta_{A\uparrow}\vert}{2}}F_{A\uparrow}(\rho)-i[2e^{- \rho /2}\rho^{\frac{\vert \zeta_{B\downarrow}\vert}{2}}(\sqrt{\frac{B_{0}}{2\rho}}F_{B\downarrow}(\rho)\vert \nu_{+}+\frac{1}{2}\vert +\sqrt{\frac{B_{0}\rho}{2}} \partial_{\rho}F_{B\downarrow}(\rho))]}{\epsilon +\triangle  }\\ \nonumber \\
\frac{-p\triangle  e^{-\rho /2}\rho^{\frac{\vert \zeta_{B\downarrow}\vert}{2}}F_{B\downarrow}(\rho)-i[2e^{- \rho /2}\rho^{\frac{\vert \zeta_{A\uparrow}\vert}{2}}(-\sqrt{\frac{B_{0}\rho}{2}}F_{A\uparrow}(\rho)+\sqrt{\frac{B_{0}\rho}{2}} \partial_{\rho}F_{A\uparrow}(\rho)])}{\epsilon +\triangle  }\\ \nonumber \\
e^{-\rho/2}\rho^{\frac{\vert\zeta_{B\downarrow}\vert}{2}}F_{B\downarrow}(\rho)\\ \nonumber \\
\frac{-p\triangle  e^{-\rho /2}\rho^{\frac{\vert \zeta_{A'\downarrow}\vert}{2}}F_{A'\downarrow}(\rho)+i[2e^{- \rho /2}\rho^{\frac{\vert \zeta_{B'\uparrow}\vert}{2}}(-\sqrt{\frac{B_{0}\rho}{2}} \partial_{\rho}F_{B'\uparrow}(\rho)])}{\epsilon +\triangle  }\\ \nonumber \\
e^{-\rho/2}\rho^{\frac{\vert\zeta_{A'\downarrow}\vert}{2}}F_{A'\downarrow}(\rho)\\ \nonumber \\
e^{-\rho/2}\rho^{\frac{\vert\zeta_{B'\uparrow}\vert}{2}}F_{B'\uparrow}(\rho)\\ \nonumber \\
\frac{p\triangle  e^{-\rho /2}\rho^{\frac{\vert \zeta_{B'\uparrow}\vert}{2}}F_{B'\uparrow}(\rho)+i[2e^{- \rho /2}\rho^{\frac{\vert \zeta_{A'\downarrow}\vert}{2}}(\sqrt{\frac{B_{0}}{2\rho}}F_{A'\downarrow}(\rho)\vert \nu_{-}-\frac{1}{2}\vert + \sqrt{\frac{B_{0}\rho}{2}} \partial_{\rho}F_{A'\downarrow}(\rho)])}{\epsilon +\triangle  }
\end{array}
\right]\nonumber 
\end{eqnarray}
for the seeking eigenfunctions (eight-component spinor) in the presence of the intrinsic SOC and a single wedge disclination. Here, $C_{n,j}$ is the normalization constant. 

The corresponding eigenvalues (Landau levels) are given by
\begin{eqnarray}\label{landaulevels}
\epsilon_n = \sqrt{4 \omega \left[n +\frac{1}{2} \left \vert \nu_{\tau}- \frac{\sigma_{z}}{2} \right \vert +\frac{1}{2} \left(\nu_{\tau}-\frac{\sigma_{z}}{2}\right) + \frac{1}{2}(1 + s_{z}) \right] + (p^{2}+h^{2})\triangle^{2}}.
\end{eqnarray}

Notice that both the spinor wave functions $\psi(\rho,\theta)$ and the energy levels $\epsilon_{n,N}$ depend on the disclination parameter $\alpha_{N}$  as well as on the modified intrinsic SOC parameter $\sqrt{ (p^{2}+h^{2})}\Delta$ and applied magnetic field $B_0$.  The solutions for Landau levels in absence of defects are contained in this solutions.  Note that, all solution of Eq.(25) and Eq.(26) are labeled $\alpha_{N}$,  the subset where $\alpha_{N}=1$ represents the Landau levels solutions for flat graphene, for case where $\alpha_{N}\neq 1$ we have all solution for Landau levels in Kane-Mele model for  disclinated graphene.  It can be verified that in the limit when $\alpha_{N}=1$ (which corresponds to $N=0$), we recover the results for the Landau levels for the  Kane-Mele model with intrinsic SOC contribution in the absence of disclinations.
\begin{figure}
\centerline{\includegraphics[scale=0.77]{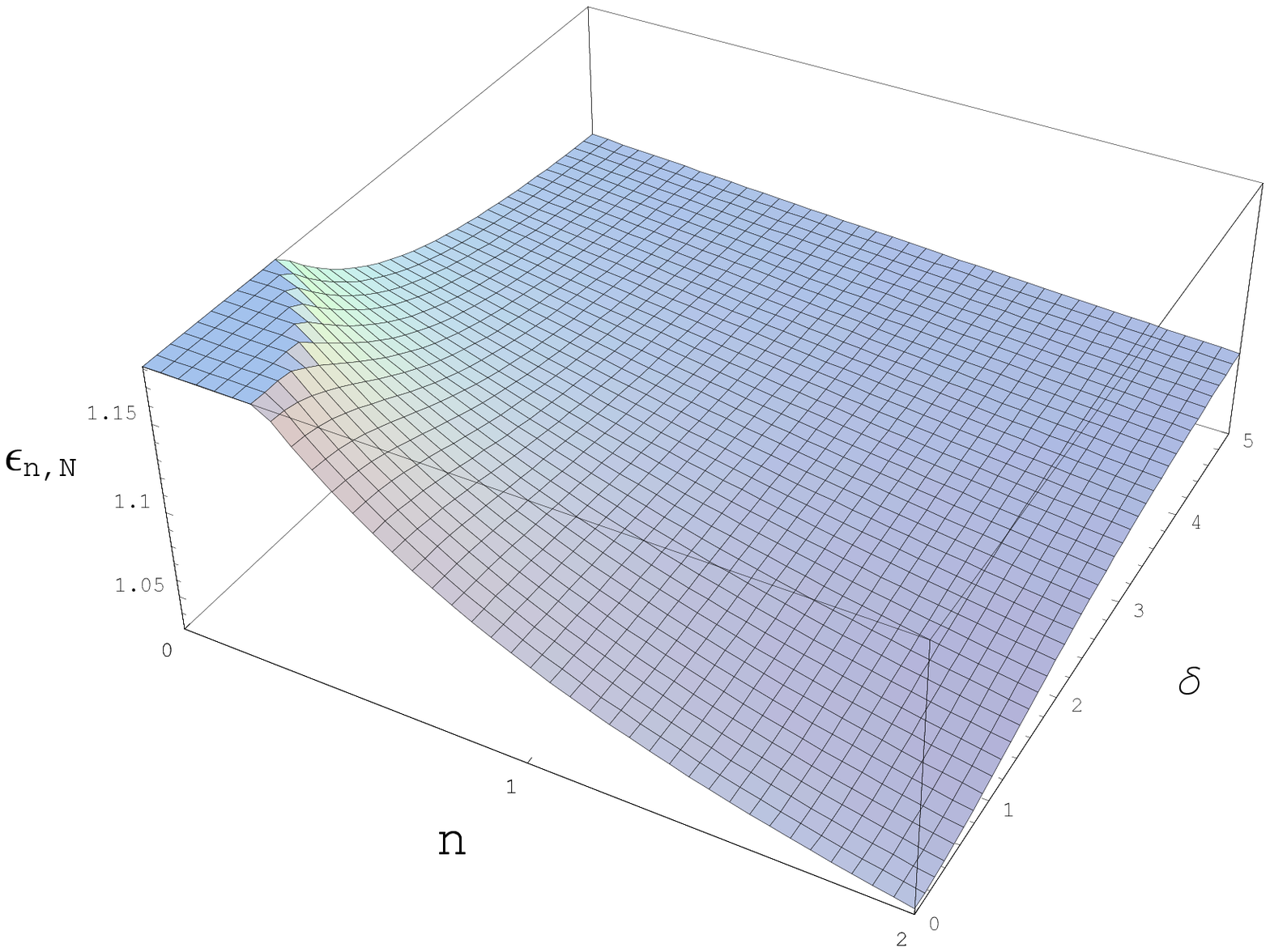}  }\vspace{0.05cm}
\centerline{\includegraphics[scale=0.77]{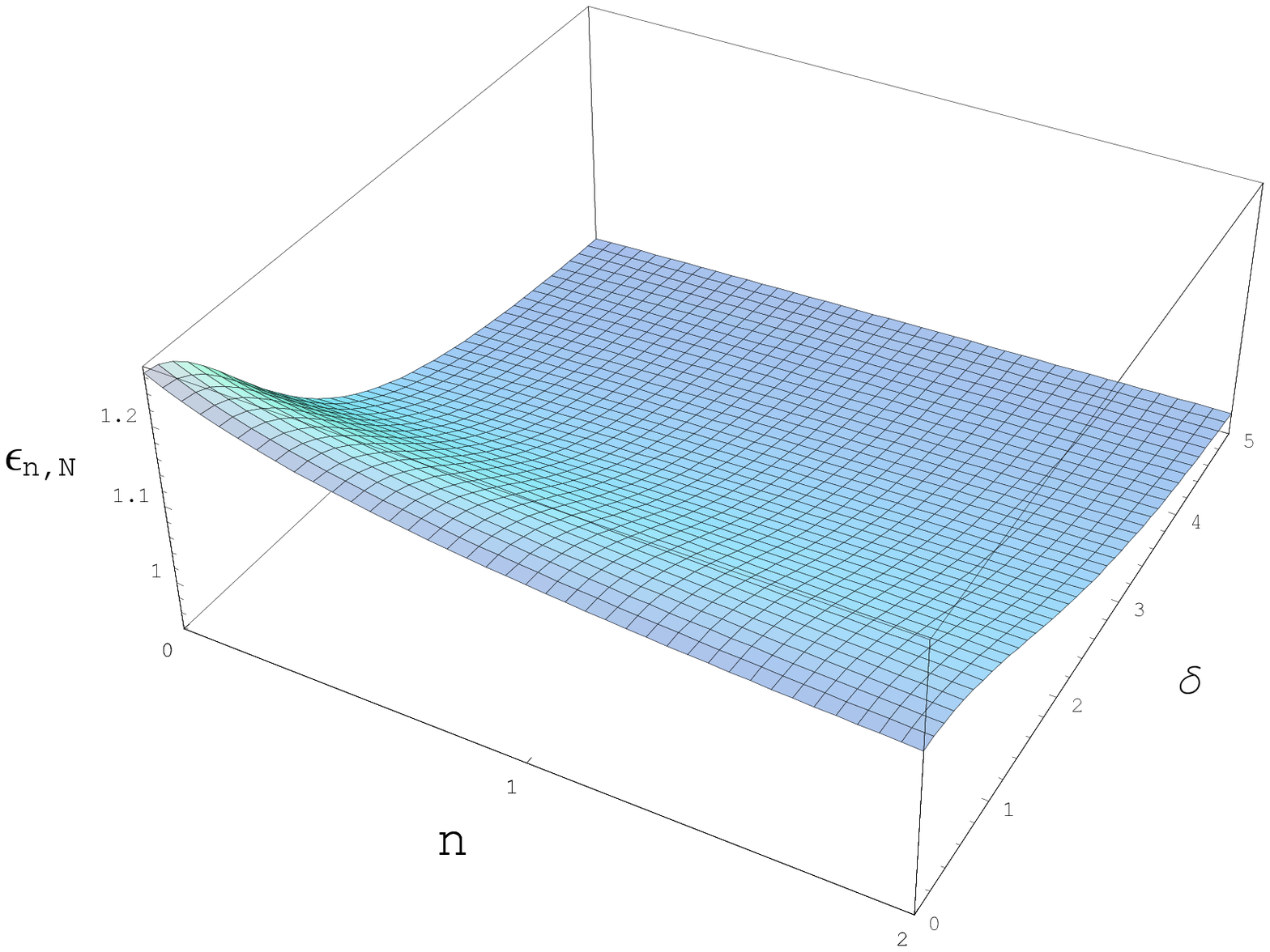}  }\vspace{0.05cm}
\caption{(Color online) Three-dimensional graphics for the evolution of the Landau levels $\epsilon_{n,N}$ (normalized to the defect-free energy $\epsilon_{n,0}$) with the quantum number $n$ and the SOC-to-magnetic field ratio $\delta$ in the presence of a positive disclination with $N =+1$ (top) and a negative disclination with $N =-1$ (bottom). }
\label{fig:fig2}
\end{figure}
Fig.2 illustrates the defects mediated evolution of the Landau levels $\epsilon_{n,N}$ (normalized to the defect-free energy $\epsilon_{n,0}$) with the quantum number $n$ for different values of the ratio $\delta =(\Delta l_B/\hbar v_f)^2$ (responsible for the competition between the SOC contribution $\Delta$ and external magnetic field $B_0$ defined via the magnetic length $l_B=\sqrt{\hbar/eB_0}$). 
\begin{figure}
\centerline{\includegraphics[scale=0.77]{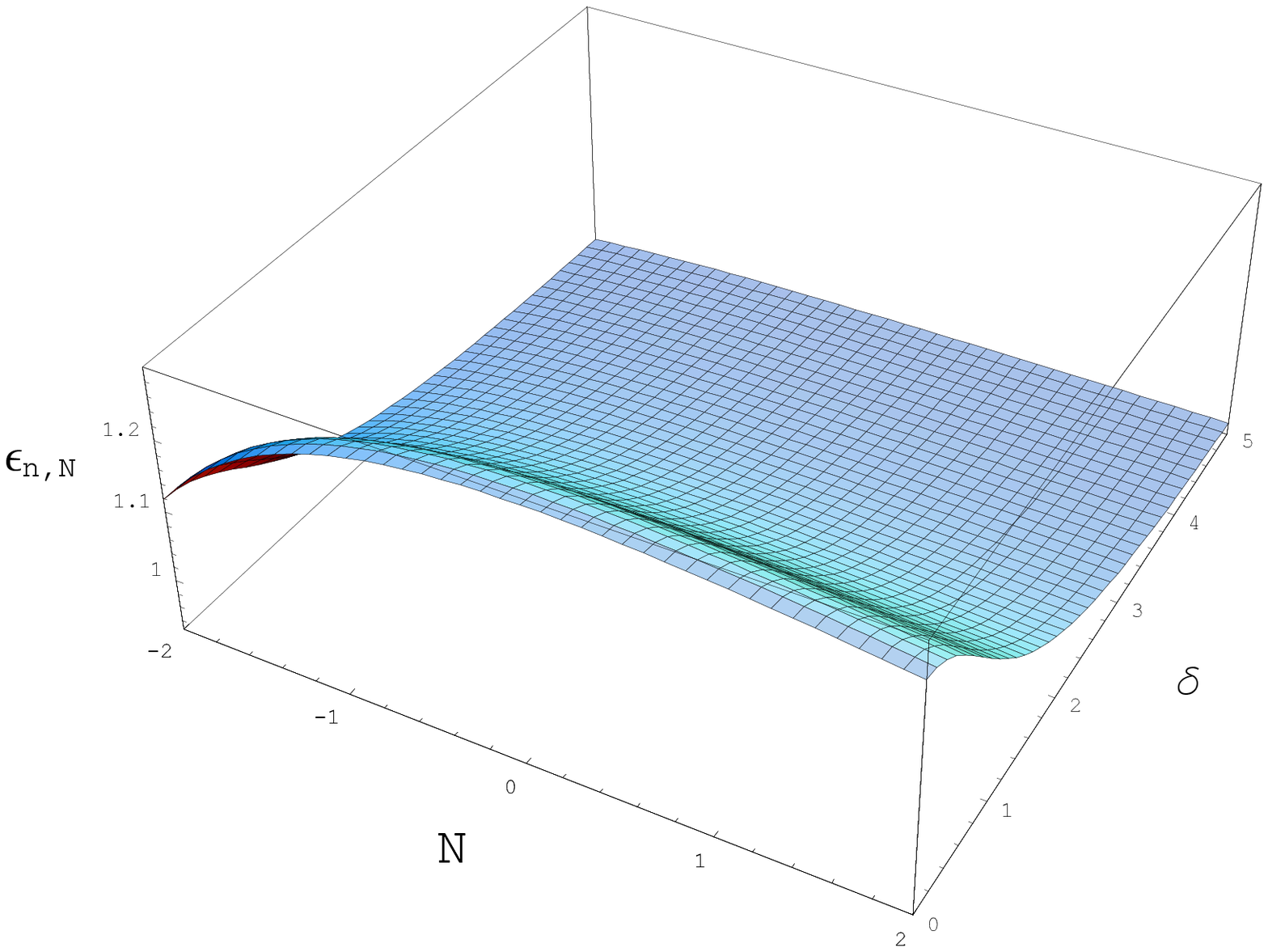}  }\vspace{0.05cm}
\caption{(Color online) Three-dimensional graphics for the evolution of the normalized Landau levels $\epsilon_{n,N}$ with the number of disclinations $N$ and the SOC related parameter $\delta$ for the ground state $n=0$.}
\label{fig:fig3}
\end{figure}
In turn, Fig.3 depicts the dependence of $\epsilon_{n,N}$ on $\delta$ and the  disclination number $N$ for the ground state $n=0$. 
For preparing these figures, the following model parameters were used: $\alpha_{N}=1 + \frac{N}{6}$, $\sin\beta=1- \frac{N}{6}$, $\tau=+1$, $\sigma_z=+1$, $s_z=+1$, $j=+\frac{1}{2}$, and $\Phi=0$. Notice that according to Fig.3, in the presence of wedge disclinations, the ground state markedly evolves with $N$ and the SOC related ratio $\delta$.

\section{Conclusions}\label{sec4}
By introducing a modified Kane-Mele Hamiltonian in the presence of a single wedge disclination, we presented exact  results on the eigenvalues and eigenvectors for the modified Landau levels in a single layer graphene with   intrinsic spin orbit coupling (SOC). We found that the presence of disclinations  changes the intrinsic SOC  which now depends on the parameter $\alpha_{N}$.  We also observed that the eigenvalues of the energy are dependent on parameter of disclination and that the presence of defect breaks the degeneracy of energy levels. In addition to fictitious magnetic flux introduced by the presence of topological defect in honeycomb lattice due to the pseudospin and given by the shift $\frac{N\tau}{4}$ in the Hamiltonian, the presence of the Aharonov-Bohm flux also contributes to the shift of the qunatum number $j$.  Furthermore, Fig.2 demonstrates that the curvature introduced by topological defect changes the behavior of the Landau levels energy $\epsilon_{n,N}$ which decreases with quantum number $n$ for positive disclination and increases with the parameter $\delta$. In the case where the topological defect introduces negative curvature in topological insulator (due to negative disclination $N=-1$), $\epsilon_{n,N}$  decreases more slowly with $n$ but its behavior as a function of $\delta$ is inverse to the previous case and the energy decreases with variation of the SOC-to-magnetic field ratio. Other interesting feature is the predicted behavior of the ground state of Landau levels $n=0$ as a function of a disclination type (given by disclination number $N$) shown in Fig.3. More precisely, we observed that the ground state is sensitive to the presence of curvature introduced by topological defect, decreasing with parameter $\delta$. Based on our findings, we may conclude that the considered here model can be used to study the influence of topological defects on quantum Hall effect in topological insulators.

{\bf Acknowledgements}
We thank CAPES, CNPQ, and FAPESQ for financial support.


\end{document}